\documentclass[11pt]{article}

\usepackage[margin=1in]{geometry}
\usepackage{amsmath,amssymb}
\usepackage{booktabs}
\usepackage{caption}
\usepackage{enumitem}
\usepackage{graphicx}
\usepackage{microtype}
\usepackage[numbers,sort&compress]{natbib}
\usepackage{tabularx}
\usepackage{tikz}
\usepackage{xcolor}
\usepackage{xurl}
\usepackage[hidelinks]{hyperref}

\usepackage[title]{appendix}
\usepackage{float}

\usetikzlibrary{arrows.meta}

\newcommand{\Rstar}{R^{*}}
\newcommand{\Kstar}{K^{*}}
\newcommand{\Tstart}{T_{\mathrm{start}}}
\newcommand{\TA}{T_{A}}
\newcommand{\TB}{T_{B}}

\newcommand{\cvar}{\mathrm{CVaR}}

\title{A Declining CVaR Glidepath Framework for Target-Date Fund Design with an Application to the Chilean Pension System}
\author{%
Israel Muñoz\textsuperscript{b}, Fernando Suárez\textsuperscript{a}, Omar Larré\textsuperscript{a}, Arturo Cifuentes\textsuperscript{b}\\[0.5em]
\small \textsuperscript{a} Fintual Administradora General de Fondos S.A. Santiago, Chile. Fintual, Inc.\\
\small \textsuperscript{b} Clapes UC, Pontificia Universidad Católica de Chile, Santiago, Chile.\\
\small Contact email: \href{mailto:research@fintual.com}{research@fintual.com}
}
\date{September 22, 2026}

\begin{document}

\maketitle

\begin{abstract}
We propose a general framework for designing Target-Date Funds (TDFs) based on a 
clear return objective, and controlling risk at the portfolio level through a 
declining Conditional Value-at-Risk (CVaR) constraint,  rather than through 
age-dependent asset-class limits or bands. In our approach, designing the 
TDF reduces to specifying a CVaR glidepath that leaves a portfolio manager 
sufficient flexibility to achieve a desired return with a reasonably high 
probability. The target return is specified  exogenously---determined  by 
factors such as retirement age, contribution rate, years worked, life 
expectancy, and a target replacement rate---and the glidepath is 
calibrated so that the risk assumed at each point in time is commensurate 
with achieving it. This departs significantly from the conventional TDF 
architecture, in which age-dependent asset-class limits are specified 
without reference to any explicit return objective.

A key methodological innovation concerns the modeling of the portfolio 
manager's behavior. Previous approaches typically assume that the manager 
succeeds every period in selecting an optimal portfolio---a strong and 
often unrealistic assumption. By contrast, we only require that the 
manager select, each month, an asset allocation drawn at random from the 
set of portfolios that satisfy the CVaR constraint imposed by the 
glidepath. This conservative assumption yields a more realistic evaluation 
of each candidate glidepath: rather than a best-case bound, the success 
probability reflects an average over the admissible allocation space, 
independent of the manager's ability to identify an optimal portfolio. 
To compare competing glidepaths, we introduce two figures of merit that 
capture, respectively, the probability of meeting the return target and 
the cumulative risk assumed over the TDF's life.

As a proof of concept, we apply the framework to Chile's 2025 pension 
reform, using nine Chilean and global asset classes and a 40-year 
accumulation horizon. Two findings stand out. First, the transition 
age---the age at which the fund begins reducing risk---is the single most 
consequential design parameter: glidepaths that begin de-risking too early 
consistently fail to deliver the required return, regardless of how 
aggressively risk is taken in the early phase. Second, contribution 
density acts as a hard constraint: below a critical threshold, no 
glidepath succeeds, indicating that portfolio design alone cannot 
compensate for structurally low contribution rates. The framework is 
general and applies to any TDF designed around an explicit return 
objective.

\end{abstract}

\noindent\textbf{Keywords:} Target-Date Funds; CVaR Glidepath;  Pension Funds; Lifecycle Investing; Replacement Rate; Contribution Density; Chilean Pension System

\section{Introduction}

In recent decades, the most significant change in the global pension landscape has been the progressive shift from defined-benefit to defined-contribution plans. Within the defined-contribution universe, Target-Date Funds (TDFs)—also known as generational funds, lifecycle funds, or age-based funds—have emerged as increasingly important vehicles. At the end of 2025, TDFs managed almost \$5 trillion, whereas in 2000 the size of this market was less than  \$8 billion \cite{roy2026,parker2023}.

Broadly speaking, a TDF is a diversified, multi-asset investment vehicle designed to serve as the primary---and often sole---retirement savings instrument for an individual with a known or anticipated retirement date. Although in principle such funds could invest in a wide variety of assets, most TDFs limit themselves to stocks and bonds (domestic and international) and have minimal or zero exposures to real estate, currencies, commodities and alternative assets. TDFs can also follow an active or passive strategy.

At present, early 2026, Vanguard is the largest  player in this space: with \$1.8 trillion in AUM, it accounts for 37\% of this segment. Fidelity is a distant second with \$693 billion. All in all, the five biggest players account for roughly 80\% of this market \cite{roy2026}.

The rationale behind most of these funds is that younger workers are better prepared to take more risk while workers approaching retirement should be exposed to safer assets. Risk, in short, should follow a smooth, time-dependent declining pattern---predefined at inception---known as the glidepath. Moreover, the conventional wisdom is that stocks are riskier than bonds, and thus, the asset composition of a TDF is expected to start with a high exposure to stocks, which should decline over time; the allocation to bonds, on the other hand, should increase over time, following a pattern opposite to that of stocks. For example, the Vanguard TDF VSVNX (target date, 2070), as of March 31, 2026, had 9.7\% in bonds. In contrast, the Vanguard TDF VTTVX (target date, 2025) had 51.3\% in bonds \cite{vanguard2070,vanguard2025}.

Three features of this conventional architecture deserve special 
attention. The first is the premise that risk in a TDF should decline 
as workers age. We accept this premise without further argument. The 
intuition is straightforward: older workers have less time to recover 
from a major portfolio loss and therefore should be exposed to 
progressively lower risk as they approach retirement. We note, in 
passing, that the classical lifecycle models---Samuelson 
\cite{samuelson1969}, Merton \cite{merton1969}, and the human-capital 
extensions \cite{bodie1992,gomes2008,cocco2005}---provide only 
ambiguous support for this convention.  Moreover, they rely  on utility-theoretic 
primitives---preference specifications and utility functions---whose 
empirical relevance remains contested. We therefore do not anchor our 
framework to those models.

Second, most TDFs rest implicitly on a strong assumption: that risk 
is an intrinsic, time-independent property of each asset class---that 
stocks are simply ``risky'' and bonds simply ``safe,'' independent of 
the portfolio in which they sit, of cross-asset correlations, or of 
the prevailing market regime. We challenge this assumption and argue 
that it is more sensible to manage risk \emph{directly}, through a 
portfolio-level metric such as the CVaR, rather than \emph{indirectly}, 
through time-dependent asset-class limits or bands.

Third, conventional TDFs do not aim, explicitly or otherwise, at a minimum return (or minimum capital accumulation level). In this sense, TDFs differ markedly from other long-term investment vehicles---endowments, for example---which typically have well-defined return targets and specific risk limits (e.g., maximum drawdowns, CVaR limits).

In this study, we depart from the usual TDF architecture and instead focus on TDFs with well-defined return targets whose risk is managed through a portfolio-level risk metric, rather than asset allocation limits. To be clear, no investment can guarantee a minimum return. The idea, however, is to structure the TDF in such a way that the level of risk it assumes remains commensurate with achieving its target return with high probability. All of these terms, of course, require precise definitions---a topic we address in detail later in this document.

 The motivation for this study, and for the innovations outlined above, 
is the recent pension reform enacted in Chile \cite{ley21735}. The 
centerpiece of this reform is the creation of a TDF-based pension 
system that will fully replace the previous regime---a framework built 
around five funds known as A, B, C, D, and E, each carrying a distinct 
risk profile, ranging from Fund A, the highest-risk, equity-oriented 
option, to Fund E, the most conservative, fixed-income-oriented option. Under the new scheme, enrollment will be mandatory, and workers will be automatically assigned to a TDF based on their age bracket, without regard to any other consideration (e.g., desired retirement age, additional assets, or outside savings). The only choice available to workers will be the selection of the asset manager overseeing the TDF---known locally as a \emph{fondo generacional}---in which they are enrolled. As of this writing, the Chilean reform is still awaiting implementation: the new investment regime for the  \emph{fondos generacionales}  must be issued by September 2026, and the transition from the current \textit{multifondos} system is expected in April 2027 \cite{previsionsocial2025mercer}.

With that as background, the goal of this effort is twofold. First, we present a general method for designing the glidepath of a TDF using a portfolio-level risk metric---specifically, a declining CVaR curve---calibrated to achieve a desired target return. In short, the glidepath is constructed so that the level of risk assumed at each point in time is consistent with the return the TDF seeks to deliver. Second, as a proof of concept, we apply this method to the Chilean pension reform. We show that under realistic conditions it is possible to structure TDFs such that the resulting pensions deliver satisfactory replacement rates with high probability. It should be noted, however, that the approach presented is entirely general and can be applied to any TDF; it is not restricted to the particular features of the Chilean reform.

The remainder of the paper is organized as follows. The next section reviews the relevant literature, after which we formally state the problem we seek to solve and provide a detailed description of our methodology (computational strategy). We then showcase how this approach works using the Chilean case as an example of application, discuss our results and conclude with a summary of our main findings.

\section{Literature Review}

Although the idea of declining-risk lifecycle portfolios is older, 
TDFs only became dominant in practice after the Pension Protection 
Act of 2006 in the United States and the subsequent Department of 
Labor QDIA rule, which allowed target-date products to serve as 
Qualified Default Investment Alternatives in defined-contribution 
plans \cite{martin2007,dol2007}. Their growth since has been rapid and largely policy-driven, 
and the literature on TDFs has expanded accordingly.

The earliest TDF designs relied on simple age-based heuristics---most famously the ``100-minus-age'' rule, which sets the equity allocation to 100 minus the participant's age. Subsequent work has shown that such rules can produce inadequate retirement outcomes under realistic market conditions \cite{shiller2005}. A related and often overlooked distinction concerns the anchor of these rules: whether the glidepath is indexed to (i) the participant's current age, or (ii) the age at which the participant intends to retire. These two formulations can produce materially different allocations for the same individual, and the literature is not always explicit about which is in use.

Beyond the simple rules, a number of alternative glidepath designs have been proposed---fixed-mix 60/40 portfolios, rising-equity paths in retirement \cite{pfau2013}, and ``U-shaped'' trajectories \cite{estrada2019a,estrada2019b}---with mixed empirical support. None of these proposals, however, departs from the core mechanism of conventional TDFs: risk is controlled indirectly, through asset-class limits or bands.  Two examples from the industry support this view. T. Rowe Price, for instance, a company that offers investment solutions to institutional and retail investors, carried out a study that explored the merits of different TDF designs based on Monte Carlo simulations.  All the alternatives, nevertheless, were based on somewhat ad-hoc stocks-bonds combinations but with no reference to any portfolio-based risk metric \cite{trowe2016}. NISA, a company active in the investment consulting space, in a similar study, considered several stocks-and-bonds combinations which were described only qualitatively in terms of risk (that is, with no reference to any specific metric or goal) \cite{nisa2016}. 

This indirect approach has well-documented drawbacks. Empirical work 
on U.S.\ TDFs documents wide dispersion among funds with the same 
target date, reflecting unstated differences in risk and return 
assumptions \cite{elton2015,balduzzi2019}. More importantly for our 
purposes, comparative evidence from Chile and Mexico 
\cite{schlechter2019} shows that investment regulations based solely 
on asset-class limits can produce erratic fund-level risk--return 
patterns.   At present, the Chilean pension system is structured based on five  funds (known as A, B, C, D and E); Fund A is supposed to be the most aggressive and E the most conservative. However, Fund E  has outperformed 
the riskiest (A) in cumulative returns over a substantial fraction of 
rolling time-windows, and Sharpe-ratio rankings have been systematically 
inverted relative to the regulator's original intent. In Mexico, by contrast, where asset-class limits are combined with a portfolio-level VaR 
constraint, the funds rank as intended. A subsequent study, 
 building on this evidence, has argued  for replacing 
asset-class limits altogether with a single portfolio-level risk 
metric  \cite{pagnoncelli2023}.

A separate strand of the literature has advocated for adaptive 
TDFs, in which asset allocations at time $T$ depend not only on $T$ 
itself but also on accumulated wealth, market conditions, or 
realized portfolio risk \cite{forsyth2017}. Such designs have 
advantages and disadvantages.  A full discussion of this topic, however, is beyond the scope of this study. The 
framework we propose is deliberately non-adaptive: the CVaR limit 
at time $T$ is determined  \textit{ex ante}  and depends only on $T$, not on 
the participant's accumulated balance. This is a design choice, 
not an oversight.

Several authors have gone beyond describing the asset composition of TDFs and have computed outcome-based performance metrics. For instance, Basu et al. \cite{basu2011} evaluated lifecycle strategies using terminal wealth distributions, shortfall probabilities, and Expected Tail Loss; Menoncin and Vigna \cite{menoncin2017} derived ruin probabilities and efficient frontiers for DC pension schemes; and Forsyth et al. \cite{forsyth2017b} assessed TDF performance through expected terminal wealth and probability of shortfall at retirement. Yoon \cite{yoon2010} went  a step further by deriving the asset allocation from a pre-specified risk budget. However, none of these studies explicitly linked the asset composition of the TDF to the outcome metrics in the design stage.  In other words, the metrics were  used to evaluate or compare strategies \emph{ex post}, rather than to construct the glidepath itself.

A recent study by \citet{perchet2025} shares
our motivation.  These authors manage risk through a portfolio-level metric,
using a Value-at-Risk constraint within a two-step recursive optimization
framework. However, their approach differs from ours in three important
respects: it lacks an explicit return objective; 
it restricts the eligible asset universe to equities, bonds, and cash; and,
most importantly, it assumes the portfolio manager always selects the
optimal portfolio on the efficient frontier---a strong and, in our view,
unrealistic assumption. Interestingly, these same authors acknowledge that
glidepath design in TDFs is typically poorly motivated and lacks standardization
across the industry---an observation that reinforces the motivation behind
the present study.

A comprehensive and up-to-date review of the TDF literature, with many references to the Chilean pension system,  is provided in \cite{suarez2025}.

In summary, the literature offers extensive discussions of how equity 
and fixed-income exposures should evolve over the life of a TDF. 
Yet comparatively little attention has been paid to two questions 
that are, in our view, logically prior: first, what objective the 
TDF should seek to achieve---a target replacement rate or a minimum 
cumulative return, for example---and second, how to translate that 
objective into a risk budget (described  through a suitable metric) that gives the fund a high probability 
of meeting it.   In short, currently there is no universally accepted method to construct a risk glidepath based on a specific, and measurable, objective. It is this gap that the present study aims to fill.

\section{Problem Statement}

We start from the standard assumption that risk in a TDF should decline as the fund approaches its target date. As noted earlier, we do not make any \emph{a priori} assumptions regarding the risk profile of the assets in which the TDF will invest (e.g., ``safe,'' ``risky''). Instead, we control risk at a portfolio level using a suitable measure---namely, the CVaR \cite{rockafellar2000}.

The objective is to determine a CVaR-limit glidepath---a  declining CVaR limit trajectory over the TDF's life---calibrated to deliver enough capital to finance an adequate annuity at the fund's closing date.

\begin{figure}[H]
  \centering
  \begin{tikzpicture}[x=0.13cm,y=0.42cm,>=Latex]
    \draw[->,thick] (0,0) -- (82,0) node[below] {   $t$};
    \draw[->,thick] (0,0) -- (0,13) node[left] {$L(t)$};
    \draw[dashed] (0,10) -- (30,10);
    \draw[dashed] (0,3) -- (70,3);
    \draw[very thick,blue!65!black] (5,10) -- (30,10) -- (70,3);
    \draw[dotted] (5,0) -- (5,10);
    \draw[dotted] (30,0) -- (30,10);
    \draw[dotted] (70,0) -- (70,3);
    \node[below] at (5,0) {$\Tstart$};
    \node[below] at (30,0) {$\TA$};
    \node[below] at (70,0) {$\TB$};
    \node[left] at (0,10) {$A$};
    \node[left] at (0,3) {$B$};
  \end{tikzpicture}
  \caption{Schematic CVaR-limit (glidepath) curve   {$L(t)$}. The TDF is constrained to an  initial CVaR  limit $A$ until $\TA$, then it linearly reduces the limit to $B$ by retirement age $\TB$.}
  \label{fig:cvar-glidepath}
\end{figure}
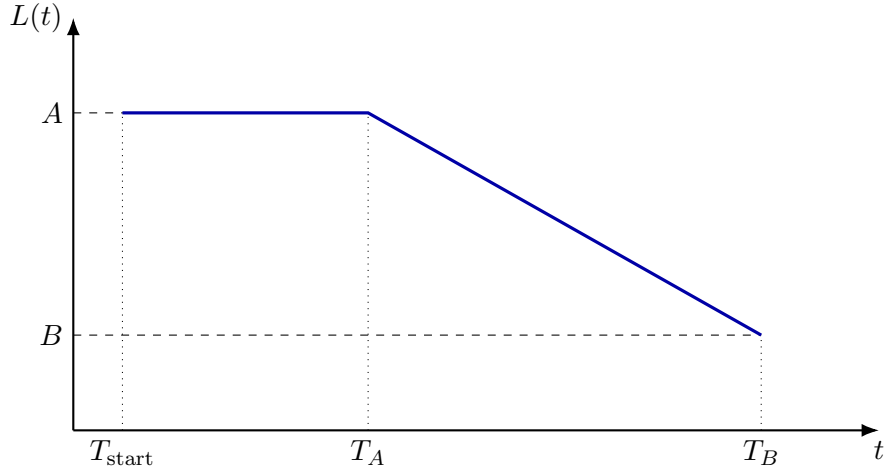

Figure~\ref{fig:cvar-glidepath} shows a typical CVaR-limit glidepath. A CVaR-limit glidepath, hereafter simply referred to as a glidepath, is defined by the following parameters: $\Tstart$, the worker's entry age into the fund; $\TA$, an intermediate age between $\Tstart$ and $\TB$; $\TB$, the worker's retirement age; and $A$ and $B$, which denote the maximum CVaR limits during the initial and final phases of the TDF's lifespan, respectively.

Formally, the CVaR limit, $L(t)$, for a time  $t$, is given by:
\begin{equation}
L(t)=
\begin{cases}
A, & \Tstart \leq t \leq \TA, \\
A + \dfrac{B-A}{\TB-\TA}(t-\TA), & \TA < t < \TB, \\
B, & t = \TB,
\end{cases}
\label{eq:glidepath}
\end{equation}

with $B < A$.  The lifespan of the TDF, $Q$, expressed in months
(assuming both $T_B$ and $T_{\text{start}}$ are expressed in years), is

\begin{equation}
Q = 12    (\TB - \Tstart).
\end{equation}

Notice that although the CVaR limit declines over time, the decline does not need to be strictly monotonic: we allow for an initial period of constant CVaR limit. Also implicit in this structure is the assumption that the TDF pools individuals from the same age cohort, who enter and exit the fund simultaneously. In other words, we assume homogeneous fund participants. We denote the minimum annualized return threshold
that the TDF needs to achieve by $R^*$.

The required return $\Rstar$ deserves careful consideration. As an exogenous input to our analysis, its value is determined by a number of factors external to the problem formulation. The process begins with an assumed wage path for the typical worker over the interval $(\Tstart,\TB)$. Using the average salary over the last ten years of employment as a reference, and applying a desired replacement rate---defined as the ratio of the monthly pension to this reference salary---it is possible to calculate the required monthly pension.

This, in turn, given the worker's life expectancy and an appropriate discount rate, makes it possible to estimate the capital $\Kstar$ needed at $\TB$ to fund an annuity. The final step closes the loop: since the wage path is known, and participants contribute a fixed fraction $\lambda$ of their salary to the TDF each month---a rate set by law and uniform across participants---$\Rstar$ can be calculated as the annualized average return that equates the future value of the monthly pension contributions to $\Kstar$.

Two important observations are in order. First, conditional on a given wage path, replacement rate, worker life expectancy, monthly contribution rate, and discount rate, the determination of $\Rstar$ is entirely deterministic. No investment strategy, of course, can guarantee that the TDF will achieve this target return. The goal, rather, is to design a glidepath whose risk profile makes it possible to reach $\Rstar$ with a reasonably high probability.

Second, we assume that the TDF can invest in a universe of $N$ assets, each with a sufficiently long return history---covering $M$ months of past monthly returns---to support simulations of future returns scenarios.

In summary, given $\Tstart$, $\TB$, and $\Rstar$---presumably determined by a regulator---and a defined universe of $N$ potential investments, the objective is to propose a glidepath (more specifically, values of $A$, $B$, and $\TA$) such that the TDF stands a high probability of attaining the required return $\Rstar$.

\section{Computational Strategy}
It is worth noting that the problem we seek to address---finding a 
glidepath that results in a satisfactory pension---is an inverse problem. 
In essence, the approach to solving such a problem consists of assuming a glidepath structure 
and then estimating the probability that it delivers the required 
return $R^*$.  If such a probability is high enough, the glidepath is deemed acceptable.

Thus, for a given CVaR glidepath $L$, defined by a given set of values $A$, $B$, $\Tstart$, $\TA$ and $\TB$, we need to estimate the probability that this glidepath can deliver the required return $\Rstar$.

Additionally, we assume that we have a universe of $N$ potential assets in which the TDF can invest, and a history of $M$ monthly return vectors for such assets. Furthermore, an asset allocation is defined as a vector of $N$ weights, $\Omega = (\omega_{1},\ldots,\omega_{N})^{\top}$, where $\omega_{i} \geq 0$ for all $i$ and $\sum_{i=1}^{N}\omega_{i} = 1$. A return vector is defined as $R = (r_{1},\ldots,r_{N})^{\top}$, where $r_i$ denotes the monthly return of asset $i$.

With that as background, we proceed as follows.

\paragraph{Simulation of returns.}
Using a Gaussian copula-based technique in combination with historical return data, we build a return-generation engine \cite{nelsen2006}. Let $S$ denote the number of simulated returns scenarios, chosen to be sufficiently large, and let $Q$ denote the number of monthly periods in the simulation horizon. The engine produces an $S \times Q$ array of simulated return vectors, denoted by $\mathcal{R}=\{R_{s,t}\}$. Each row, consisting of a sequence of $Q$ monthly return vectors, defines one return scenario, while each column corresponds to one simulation month. Each element $R_{s,t}$, with $s \in \{1,\ldots,S\}$ and $t \in \{1,\ldots,Q\}$, is the vector of simulated asset returns in month $t$ under scenario $s$.

We are mindful that many approaches exist for generating plausible returns from historical data, of which the Gaussian copula is one. Discussing their merits is beyond the scope of this study. The approach that follows is agnostic to any particular returns-generation engine.

Note that this array of simulated returns is generated once and held fixed across all candidate glidepath evaluations (a topic discussed in an upcoming subsection); reusing the same returns scenarios ensures that differences across candidate glidepaths reflect differences in the glidepath structure alone, rather than simulation noise.

\paragraph{Feasible asset allocation.}
An asset allocation $\Omega$ is deemed feasible for month $t$, with respect to the CVaR limit $L(t)$, if its estimated CVaR, computed using the $S$ simulated return vectors corresponding to that month, does not exceed $L(t)$. To calculate this CVaR, we first compute the corresponding $S$ scalar portfolio returns:
\begin{equation}
\delta_{s,t} = \Omega^{\top}R_{s,t}, \qquad s=1,\ldots,S.
\end{equation}
These scalar returns are then sorted from lowest to highest, and the CVaR is estimated as the average of the worst ten percent of them, since we work with a 90\% confidence level.  The CVaR, strictly speaking, is a negative number (it describes losses); however, for convenience, it is often customary to express it as a positive number.  We adopt this choice in what follows. 

\paragraph{Feasible asset-allocations array.}
Using a Monte Carlo sampling algorithm based on the Hit-and-Run method (see \hyperref[app:hit-and-run]{Appendix}; the algorithm follows \citet{smith1984}), we construct a feasible asset allocation array  $\mathbf{\Omega}$  of dimension $I \times Q$, where $I$ denotes the number of feasible portfolios to be constructed. Each entry $(i,t)$ of this array corresponds to an $N$-dimensional vector $\Omega$ that satisfies two conditions: (a) its CVaR limit is below $L(t)$, the limit specified by the glidepath for month $t$; and (b) the set of return vectors used to compute its CVaR corresponds to column $t$ of the return array $\mathcal{R}$. That is, the set of return vectors employed to calculate the CVaR for month $t$ is $\{R_{s,t}\}_{s=1}^{S}$.

\paragraph{Returns scenarios and feasible portfolios.}
We assume that the TDF manager rebalances the portfolio monthly. As mentioned before, a returns scenario $s$   is defined as a sequence of $Q$ return vectors of dimension $N$, that is, $\{R_{s,1},\ldots,R_{s,Q}\}$. From the returns array we have $S$ such scenarios, where each row $s$ of that array defines one such scenario.

We define a feasible portfolio $P_i$, for $i \in \{1,\ldots,I\}$, as a sequence of $Q$ feasible asset allocation vectors, that is, $P_i=\{\Omega_{i,1},\ldots,\Omega_{i,Q}\}$. To construct $I$ feasible portfolios from the asset-allocation array, rather than taking its rows directly as generated by the Hit-and-Run sampler---which could introduce spurious serial correlation between consecutive monthly asset allocations, since the Hit-and-Run algorithm generates samples sequentially and adjacent draws are not fully independent---we first apply an independent random permutation scheme within each column of this array. In short, feasible portfolio $i$ is obtained by taking the $i$th row of the  permuted array.

This construction sets the stage for modeling the behavior of a portfolio manager who does not follow any  intertemporal strategic rule, but rather  chooses randomly, every month, an asset allocation from the set of feasible asset allocations (namely, asset allocations that satisfy the CVaR limit for that specific month).

\paragraph{Cumulative return.}
To evaluate the performance of a glidepath, each feasible portfolio $i \in \{1,\ldots,I\}$ is evaluated under each returns scenario $s \in \{1,\ldots,S\}$. The return of portfolio $i$ in month $t$ under scenario $s$ is:
\begin{equation}
r_{i,s,t} = \Omega_{i,t}^{\top}R_{s,t}.
\end{equation}
Thus the cumulative return of feasible portfolio $i$ under scenario $s$ over the life of the TDF is:
\begin{equation}
\pi_{i,s} = \prod_{t=1}^{Q}\left(1+r_{i,s,t}\right)-1,
\end{equation}
which annualized can be expressed as:
\begin{equation}
\widehat{R}_{i,s} = \left(1+\pi_{i,s}\right)^{12/Q}-1.
\end{equation}

\paragraph{Feasible glidepaths and their success probability.}
Having computed $\widehat{R}_{i,s}$ for each of the $I \times S$ (feasible portfolios $\times$ returns scenarios) combinations, we estimate $\Psi$, the fraction of cases in which $\widehat{R}_{i,s} \geq \Rstar$:
\begin{equation}
\Psi = \frac{1}{I \times S}\sum_{s=1}^{S}\sum_{i=1}^{I}
\mathbf{1}\left[\widehat{R}_{i,s} \geq \Rstar \right].
\label{eq:psi-estimator}
\end{equation}
This figure of merit can be interpreted as the probability that a hypothetical portfolio manager, who randomly selects a feasible portfolio from the set of all feasible portfolios, will succeed in delivering the required return $\Rstar$. This is a diagnostic interpretation of the feasible set, not the estimated success probability of an average professional manager. The lower the value of $\Psi$, the less desirable the glidepath is, from a regulatory standpoint.

We should note, however, that it is possible to find several glidepaths whose $\Psi$ values are sufficiently high to consider them attractive, at least in principle. This motivates incorporating an additional criterion for comparing different  glidepaths.  A glidepath that achieves a high $\Psi$ at the expense of taking excessive risk, for example, might be less attractive from a regulatory viewpoint. Thus, we can incorporate a second figure of merit, $\Gamma$, defined as follows:
\begin{equation}
\Gamma = \sum_{t=1}^{Q}L(t),
\end{equation}
whose aim is to capture the ``cumulative risk'' associated with a given glidepath.

Therefore, selecting the best glidepath is really a two-objective problem: maximizing $\Psi$ while keeping $\Gamma$ under some acceptable risk level.

\section{Example of Application}

We now apply the framework developed in the previous section to a 
concrete setting: the TDFs that will form the backbone of the future 
pension system in Chile \cite{ley21735}. Our aim is twofold: first, 
to show that the proposed methodology can be operationalized with 
realistic inputs; and second, to draw substantive conclusions about 
glidepath design for the Chilean case---conclusions that, we believe, 
will inform the ongoing regulatory discussion surrounding the 
implementation of the new system.

As emphasized earlier, the methodology itself is general---Chile merely provides a convenient proof of concept. The reader interested only in the methodology can safely skip this section.

The application proceeds in four stages. First, we estimate  the 
required return $R^*$ using assumptions about wages, contribution 
rates, life expectancy, and the target replacement rate. Second, 
we specify the universe of eligible assets available to the TDF. 
Third, we define a grid of candidate glidepaths and compute, for 
each candidate, the success probability $\Psi$ and the cumulative 
risk $\Gamma$---two figures of merit that together capture the 
trade-off between the probability of meeting the return target and 
the total risk assumed over the TDF's life. Finally, we examine 
the sensitivity of the results to two parameters of particular 
relevance in the Chilean context: contribution density and the 
age where de-risking begins ($\TA$).

\subsection{Assumptions and Basic Calculations}

\paragraph{A note on units.}
All monetary units are expressed in \textit{Unidades de Fomento} 
(UF), an inflation-indexed Chilean unit of account \cite{bcchUf}. 
For readability, we drop the ``UF'' qualifier throughout.
 Because the UF captures  inflation, every figure here should be read as a real  (inflation-adjusted) figure. 

\paragraph{Salary evolution.}
We assume a deterministic salary path consistent with the OECD methodology used in \emph{Pensions at a Glance}: constant real annual growth of 1.25\% throughout the working career \cite{oecd2025}. Starting from an index value of 100 at labor-market entry ($\Tstart=25$ years), the salary reaches a value of approximately 164 by the statutory retirement age ($\TB=65$ years).

\paragraph{Replacement rate.}
The target replacement rate $\rho$ is the ratio of the monthly pension to a reference salary at retirement. We adopt $\rho=63\%$, the OECD average net replacement rate for full-career average-wage workers reported in \emph{Pensions at a Glance} (2025) \cite{oecd2025}. For the reference salary we follow the standard pension-calculation convention and use the average salary over the final 120 months (10 years) of the working career. This window is long enough to smooth out short-run fluctuations near retirement but short enough to anchor the pension to the worker's late-career earnings rather than the lifetime average.

\paragraph{A representative worker.}
We treat the representative worker as a gender-neutral individual. 
This is appropriate for the methodological focus of the paper; the 
gender-disaggregated case---which involves different life 
expectancies, contribution densities, and statutory retirement 
ages---introduces distinct issues that we return to in 
Section  5.2.

\paragraph{Parameters specific to the Chilean case.}
The Chilean institutional setting determines five additional inputs.

The working horizon runs from $\Tstart=25$ years to $\TB=65$ years: the OECD convention sets the entry at age 22, but the average age of new affiliates in Chile is 28, thus we use the midpoint \cite{oecd2025,sp2024ficha}; $\TB=65$ years is the current statutory retirement age for men \cite{dl3500}. 
We take life expectancy at retirement to be  88 years, which according to 
the Chilean mortality-table framework published by the 
\textit{Superintendencia de Pensiones} and the \textit{Comisi\'{o}n 
para el Mercado Financiero} \cite{sp2023mortality}, represents 
the average of men (86 years) and women (90 years).
The contribution rate used in this exercise is $\lambda=16\%$ of salary, intended to capture the higher (compared to the current $10\%$) figure specified by the upcoming reform. The annuity discount rate is  3.2\% per year, in line with recent rates observed in the Chilean \emph{rentas vitalicias} (annuities) market  \cite{spRentasVitalicias}.

\paragraph{Calculation of $\Kstar$.}
$K^*$ denotes the capital the worker must have accumulated 
at retirement in order to fund a constant monthly pension 
equal to $\rho S_{\mathrm{ref}}$ over the post-retirement horizon.
It is given by:
\begin{equation}
\Kstar = \rho S_{\mathrm{ref}}a_{n|r},
\end{equation}
where $S_{\mathrm{ref}}$ is the reference salary, defined as the average  salary over the final 120 months; $n$ is the post-retirement horizon in months; $r$ is the monthly discount rate (0.00267 based on an annual discounr rate of 3.2\%); and $a_{n|r}$ is the standard present-value annuity factor.

\begin{align}
n &= 12 \times (\text{life expectancy} - \TB) = 276 \text{ months}, \\
r &= (1+0.032)^{1/12}-1 \approx 0.267\% \text{ per month}, \\
a_{n|r} &= \frac{1-(1+r)^{-n}}{r} \approx 195.
\end{align}
With these values, the target capital can be expressed as:
\begin{equation}
\Kstar \approx \rho S_{\mathrm{ref}} \times 195.
\end{equation}
Under the baseline assumptions, and assuming an initial salary 
of 20 UF per month  (roughly equivalent to USD~900, and 
consistent with the median Chilean salary), we obtain 
$K^* \approx 3{,}800$.
\paragraph{Calculation of $\Rstar$.}
Given $\Kstar$, $\Rstar$ is the constant annualized real return that, applied to the stream of monthly contributions over the working career, exactly accumulates $\Kstar$ at retirement. It is determined by first solving for the monthly rate $r^*$:
 
\begin{equation}
K^* = \lambda \sum_{t=1}^{Q} S_t (1 + r^*)^{Q-t},
\end{equation}
where $Q = 12 \times (T_B - T_{\text{start}}) = 480$ months, $S_t$ is 
the real salary in month $t$ along the path defined in the 
salary-evolution subsection, $\lambda = 16\%$, and $r^*$ is the monthly 
required rate necessary to achieve a capital of $K^*$.
The equation is solved numerically with an iterative procedure. The annualized rate is then obtained as $\Rstar = (1+r^*)^{12}-1$.

\paragraph{Contribution density.}
 
The computation above assumes the worker contributes every month 
from $T_{\text{start}}$ to $T_B$. In practice, this rarely happens. 
Workers move in and out of the formal labor market---spells of 
unemployment, self-employment, informal work, or caregiving 
interruptions all translate into months of missing contributions. 
These gaps---known locally as \textit{lagunas previsionales}---are 
a structural feature of the Chilean pension system and, more 
broadly, of the Latin American experience. The \textit{Superintendencia de 
Pensiones} reports average contribution densities of approximately 
$58\%$ for men and $50\%$ for women \cite{sp2024ficha}.
 
Modeling each worker's individual contribution pattern explicitly is beyond the scope of this study. Instead, we adopt a transparent approximation: rather than alternating between full contributions 
and zero, we assume the worker contributes a constant fraction 
$\beta$ of the statutory amount every month over the working career, 
with $0 < \beta \leq 1$. This preserves total career contributions 
while smoothing them uniformly across months.
 
Therefore, the savings equation is modified accordingly. The monthly 
contribution becomes $\beta\lambda S_t$, and $r^*$ is found by solving:
\begin{equation}
K^* = \beta\lambda \sum_{t=1}^{Q} S_t (1 + r^*)^{Q-t}.
\end{equation}
Under the baseline density $\beta=60\%$---chosen to be slightly above the observed male average of 58\%, so that the design problem is feasible while remaining realistic---and the other parameters defined previously, the solution is r* = 0.004472, which yields 

\begin{equation}
\Rstar = (1+r^*)^{12}-1 = 5.5\%.
\end{equation}

This value anchors all subsequent calculations.

\paragraph{Eligible assets.}
The standard U.S.\ TDF draws from a narrow asset universe---typically 
two broad classes, equities and bonds---with the shift between them 
driving essentially all of the time variation in risk. We take a 
deliberately broader view. Consistent with our portfolio-level risk 
framework, we make no prior assumptions about which assets are 
``risky'' and which are ``safe'': risk is controlled at the portfolio 
level via the CVaR limit, and the algorithm is free to combine any 
eligible assets so long as the resulting portfolio CVaR remains within 
the glidepath. We therefore work with a universe of $N=9$ asset 
classes---spanning Chilean fixed income, Chilean equities, global fixed 
income, global equities, and alternatives---that constitutes a realistic 
opportunity set for a Chilean pension manager. The historical record 
covers April 2007 through December 2025 at monthly frequency 
($M=225$ observations). Table~\ref{tab:asset-universe} reports summary 
statistics for each asset class; CVaR is computed at the 90\% confidence level.  Returns are expressed in Chilean inflation-adjusted (UF) pesos; no currency hedging is involved.

\begin{table}[H]
  \centering
  \small
  \caption{List of eligible assets: key statistics based on monthly real returns (inflation-adjusted Chilean pesos); time period covered, April 2007-December 2025.}
  \label{tab:asset-universe}
  \begin{tabularx}{\textwidth}{@{}>{\raggedright\arraybackslash}X l r r r@{}}
    \toprule
    Asset class (Indices) & Ticker & Mean return & Std. dev. & CVaR (90\%) \\
    \midrule
    LVA-Chile Certificates of Deposit (CD) & LKXIP & 0.01\% & 0.37\% & 0.62\% \\
    Riskamerica Corporate Bonds (Chile) & RACLCORP & 0.27\% & 1.51\% & 2.87\% \\
    S\&P/CLX IPSA (Chilean equities) & IPSA & 0.34\% & 4.92\% & 8.22\% \\
    Bloomberg Buyout Index & PEBUY & 0.84\% & 4.33\% & 6.13\% \\
    Global Aggregate (fixed income) & LEGATRUU & 0.13\% & 3.22\% & 5.10\% \\
    U.S. Corporate High Yield & LF98TRUU & 0.47\% & 3.43\% & 5.62\% \\
    MSCI World (Global equities) & NDDUWI & 0.63\% & 4.41\% & 7.78\% \\
    MSCI Emerging Markets (EM equities) & NDUEEGF & 0.39\% & 4.92\% & 8.56\% \\
    Nasdaq 100 & NDX & 1.31\% & 5.34\% & 8.25\% \\
    \bottomrule
  \end{tabularx}
\end{table}

A few remarks on the asset universe are in order. For the Chilean 
fixed-income component, we use the Riskamerica Corporate Bonds
index rather than a broader aggregate benchmark in order to isolate 
a spread-bearing corporate exposure. The index excludes the sovereign 
and base-rate component but retains corporate issuers, including 
banks. This choice is motivated by the fact that the Chilean 
bank-bond market is liquid, deep, and generally highly rated, while 
still offering a meaningful spread over government debt. A purely 
sovereign benchmark would provide a lower-yield proxy and would make 
the simulation exercise less informative. More generally, the asset 
universe is not intended to exhaust the instruments available to 
pension managers in practice---the eligible investment universe is 
substantially larger. The purpose here is to use a compact and 
representative set of proxies to showcase the methodology, rather 
than to endorse any particular choice for the investment universe.

The single-asset CVaR ranges from 0.62\% (Chilean CDs) to 8.56\% (emerging markets equities), spanning more than a factor of ten. This wide dispersion gives the algorithm real flexibility in constructing portfolios anywhere along a declining-CVaR glidepath, and it helps calibrate the search space for the glidepath parameters $A$ and $B$ (discussed in the following subsection). Historical monthly returns feed the Gaussian-copula engine described in Section 4, to generate the $S=10{,}000$ returns scenarios used throughout the analysis.

\paragraph{Candidate  glidepaths.}
We fix $\Tstart=25$ years and $\TB=65$ years throughout this example, since these are structural features of the Chilean case rather than parameters to optimize. The TDF simulation horizon is therefore $Q=\TB-\Tstart=40$ years which we express as 480 months for the purpose of this analysis. The CVaR limits $A$ and $B$ are constrained by the maximum and minimum values associated with the eligible assets (Table~\ref{tab:asset-universe}).  Note that $A$ cannot realistically  exceed the maximum single-asset CVaR (8.56\%), while $B$ cannot fall below the minimum (0.62\%). On this basis we limit our exploration to $A \in \{5\%,6\%,7\%,8\%,9\%,10\%\}$, fix $B=3\%$---a conservative but attainable floor, close to the CVaR of the Riskamerica Corporate index---and let the transition age $\TA$ (for convenience expressed in years) vary over the integer range $[30,64]$. This yields a grid of $35 \times 6 = 210$ candidate glidepaths.

\paragraph{Candidate glidepath assessment.}

 First, using the Gaussian-copula engine described in Section 4, we generate $S=10{,}000$ monthly returns scenarios over the 480-month horizon---a single $(S \times Q)$ returns array that is reused to evaluate all glidepath candidates. Then, for each of the 210 candidate glidepaths we proceed in two steps.  First, for each glidepath, we generate $I=10{,}000$ feasible portfolio trajectories---sequences of asset weights that every month satisfy the CVaR-limit constraint specified by the glidepath (by generating monthly feasible asset allocations with the Hit-and-Run algorithm described in the Appendix).  And second,  we evaluate every feasible portfolio--returns scenario pair, obtaining $I \times S=10^8$ simulated outcomes from which we compute the success probability $\Psi$ and the cumulative risk $\Gamma$ for that glidepath.

\subsection{Simulation   Results and Analyses}

Unless otherwise noted, the results use the Gaussian-copula return model,
the nine-asset universe, and the constant-fraction approximation to
contribution density described in Section 5.1. The reported thresholds
apply to the 210 candidate glidepaths considered here and the
$\Psi>50\%$ success criterion.

\begin{figure}[H]
  \centering
  \includegraphics[width=0.9\textwidth]{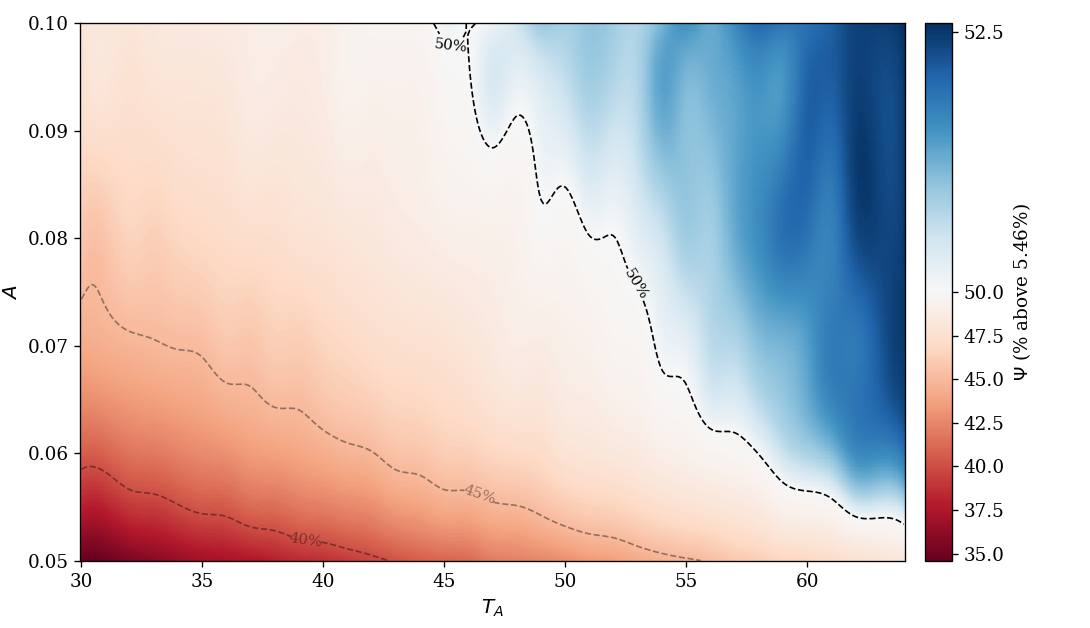}
  \caption{Heatmap of success probability $\Psi$ over the $(A,\TA)$ grid for $\Rstar=5.5\%$. The 50\% contour separates successful from unsuccessful CVaR glidepaths.}
  \label{fig:heatmap}
\end{figure}

\paragraph{Base  Case.}
Exploring the merits of each of the 210 CVaR glidepaths is essentially an exercise in 
exploring the merits of the different combinations of $A$ and $T_A$. Of the 210 candidate 
glidepaths, 66 are successful in the sense that $\Psi > 50\%$; that is, their probability 
of exceeding the target return $R^* = 5.5\%$ exceeds 50\%. Figure~\ref{fig:heatmap} displays $\Psi$ as a heatmap over the parameter grid, with the 
initial CVaR limit $A$ on one axis and the transition age $T_A$ on the other.

Three features of the heatmap are worth highlighting.
First, $\Psi$ increases with both $A$ (the maximum CVaR limit admissible during 
the initial constant-risk period) and $T_A$ (the age at which the fund begins 
reducing risk). Along each axis taken alone, $\Psi$ increases monotonically---
subject to small fluctuations attributable to simulation noise---and the two 
effects reinforce each other: allowing more risk ($A$) and sustaining it for 
longer ($T_A$) are complementary routes to clearing the $\Psi > 50\%$ threshold.

Second, the boundary between successful and unsuccessful glidepaths reveals clearly the trade-off between the two design parameters. For each value of $A$ at or above $6\%$, there is a minimum transition age below which the glidepath fails---and this minimum age decreases as $A$ increases. In other words, if the regulator is willing to allow more risk during the initial period (a higher $A$), the transition to safer assets can begin earlier; if instead the maximum risk $A$ is held to a lower level, the transition must be delayed correspondingly. Table~\ref{tab:boundary} summarizes this trade-off.  Notice also that increasing $T_A$ beyond the minimum value required to achieve success  only marginally improves the value of $\Psi$.

\begin{table}[H]
  \centering
  \caption{Minimum successful transition age $\TA$  by initial CVaR limit ($A$)  }
  \label{tab:boundary}
  \begin{tabular}{@{}lccc@{}}
    \toprule
    $A$ & Min.  $\TA$ (years) & $\Psi$ at boundary & Maximum $\Psi$ for this  $A$ \\
    \midrule
    5\% & none & -- & 48.2\% \\
    6\% & 58 & 50.0\% & 51.9\% \\
    7\% & 54 & 50.1\% & 52.5\% \\
    8\% & 53 & 50.3\% & 52.5\% \\
    9\% & 47 & 50.2\% & 52.5\% \\
    10\% & 45 & 50.1\% & 52.6\% \\
    \bottomrule
  \end{tabular}
\end{table}

Also noticeable, glidepaths with $A=5\%$ never reach the $\Psi > 50\%$  
threshold for any value of $T_A$: the maximum risk level is simply too low to 
allow the TDF to take enough risk to deliver the required return $R^*$, 
regardless of when the transition begins. Glidepaths with $A=6\%$ require 
holding the maximum risk almost to retirement ($T_A=58$), while those with 
$A=10\%$ can begin the transition more than a decade earlier ($T_A=45$). The 
two design levers---how much risk to allow and for how long---are partial 
substitutes.

Third, among all successful glidepaths, the one with minimum cumulative risk 
$\Gamma$ has parameters $A=6\%$, $B=3\%$, $T_A=58$, with $\Gamma=27.53$ and 
$\Psi=50.02\%$. This glidepath maintains the initial CVaR limit of $6\%$ for 
33 years and then reduces it linearly to $3\%$ over the final 7 years of the 
worker's career. We refer to this as the \emph{minimum-risk optimum}. It 
establishes a lower bound on the cumulative risk compatible with the 
$\Psi > 50\%$ success threshold. Whether this is the most desirable glidepath 
from a regulatory standpoint is ultimately a public-policy decision that goes 
beyond the scope of this simulation exercise.

Finally, two observations regarding the robustness of the simulation results are worth mentioning.  First, the computation of $\Psi$ for each glidepath was based on the percentage  of the $I \times S$ (feasible portfolios $\times$ returns scenarios) combinations that met the minimum return condition (R* = 5.5\%).  Alternatively, one could have defined success in terms of the minimum wealth required to achieve the desired pension, that is, 3,800 in this case.  (Notice, from the previous section, that the target return, 5.5\%,  was derived from the 3,800 figure.)  Computation of $\Psi$ based on this criterion (accumulated wealth) was consistent with the results obtained based on R* = 5.5\%.  In fact, averaging the discrepancies between the values of $\Psi$ computed with either approach over the 210 glidepaths analyzed yielded an average difference of less than 2\%; in essence, simulation noise rather than a conceptual difference with  practical implications.  Briefly, both approaches to assess glidepath suitability yielded consistent results.

Second, for computational efficiency the Gaussian copula-based engine was used only once to generate an $S \times Q$ array of simulated returns vectors (denoted as $\mathcal{R}=\{R_{s,t}\}$ as previously  indicated).  That is, the same array of returns vectors was employed to estimate the CVaRs (feasible portfolios) and returns along each returns scenario.  An \textit{ex post} (more computationally intense) analysis, based on two different arrays of returns (one to estimate the CVaRs, the second to estimate returns), both generated with the same Gaussian copula engine, yielded almost identical results.

\paragraph{Importance of $\TA$.}
To isolate the role of the transition age $\TA$, we consider a restricted version of the problem in which $A$ and $B$ are fixed. Specifically, we set $A=10\%$---essentially leaving the risk unconstrained initially, giving managers maximum flexibility in the early phase---and $B=3\%$ as before. The only question is then: when should the transition age begin?

\begin{figure}[H]
  \centering
  \includegraphics[width=0.9\textwidth]{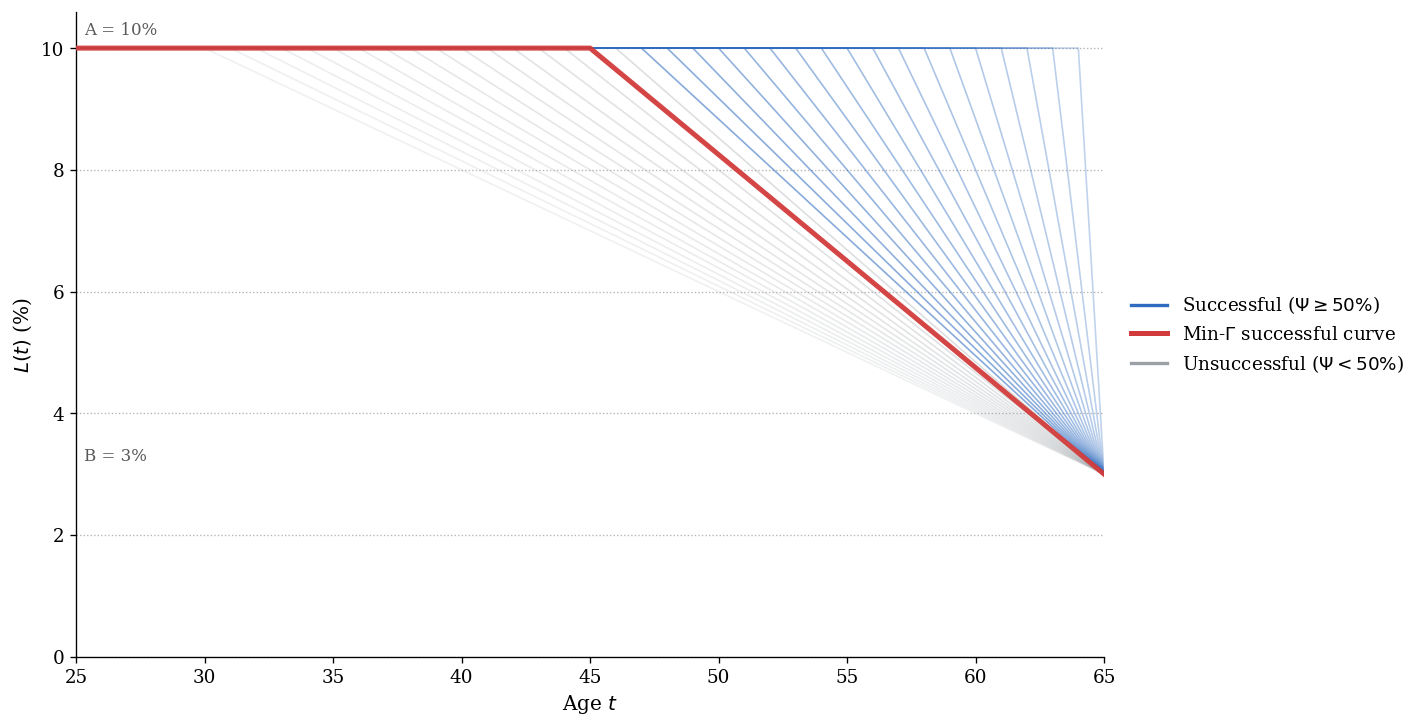}
  \caption{Filtered envelope of CVaR glidepaths for  $A=10\%$, and $B=3\%$. Successful glidepaths ($\Psi>50\%$) are shown in blue; unsuccessful glidepaths are shown in gray; and the minimum-$\Gamma$ successful glidepath  is shown in red. }
  \label{fig:filtered-envelope}
\end{figure}

The results, displayed in Figure~\ref{fig:filtered-envelope}, show that $\Psi$ 
increases monotonically with $T_A$, and first crosses the $\Psi > 50\%$ success 
threshold at $T_A=45$ ($\Psi=50.13\%$, $\Gamma=39.57$). Glidepaths that begin 
reducing risk before age 45 systematically fail to deliver the required return 
$R^*=5.5\%$, regardless of how much risk is taken during the constant-risk 
period.

This finding offers an important policy-relevant interpretation: under realistic 
Chilean parameters, the simulations suggest caution against beginning the 
downward risk transition before the worker's mid-forties. Young workers can 
afford to take more risk, as they have time to recover from adverse outcomes. 
In essence, the simulations warn against the ``risk of not taking enough risk.''

The \emph{minimum-risk optimum} serves as a useful benchmark: it identifies 
the lowest cumulative-risk glidepath compatible with the $\Psi > 50\%$ success threshold. 
The filtered parameterization ($A=10\%$, $B=3\%$, $T_A=45$) is simpler to 
communicate and more robust to changes in assumptions. Both should be read as 
inputs for regulatory discussion, not as prescriptive recommendations.

\paragraph{Importance of the contribution density.}

Chile has a well-documented problem of low and uneven contribution density, 
driven by informal employment, labor-market gaps, and gender-specific patterns. 
Since $R^*$ depends directly on the fraction of months in which the worker 
actually contributes, understanding how density affects the simulation results is essential.

\begin{figure}[H]
  \centering
  \includegraphics[width=0.9\textwidth]{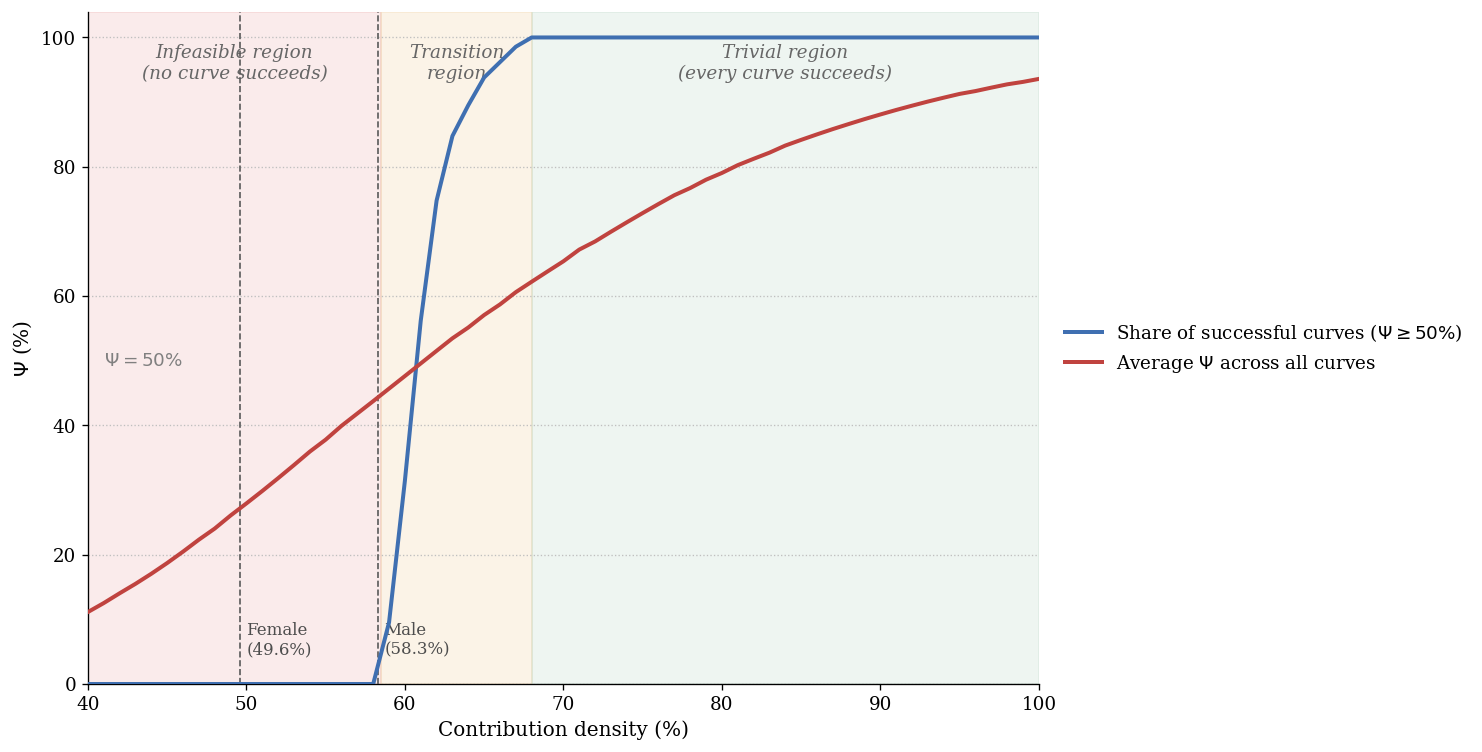}
  \caption{Percentage of successful glidepaths  ($\Psi>50\%$) and average success rate across all glidepaths, as a function of contribution density.}
  \label{fig:contribution-density}
\end{figure}

Figure~\ref{fig:contribution-density} shows two curves plotted against 
contribution density: the fraction of candidate glidepaths that clear the 
$\Psi > 50\%$ success threshold, and the average $\Psi$ across all (successful and unsuccessful) glidepaths. 
The picture reveals three distinct regions.

For densities at or below $58\%$, no glidepath in our grid achieves 
$\Psi > 50\%$: the required return $R^*$ is too high to be reached under 
any glidepath. In this regime, the problem is simply 
infeasible---no portfolio investment strategy can compensate for insufficient contributions.

For densities at or above $68\%$, every glidepath in the grid achieves 
$\Psi > 50\%$: the required return $R^*$ is low enough that even the most 
conservative glidepaths succeed.

Between these boundaries---roughly $59\%$ to $67\%$---lies the region where  glidepath design matters most. Within this window, the share of successful 
glidepaths rises steeply: $9.5\%$ at density $59\%$, $31.4\%$ at $60\%$, 
$56.2\%$ at $61\%$, and $98.6\%$ by $67\%$. This transition spans less than 
ten percentage points of density, which is remarkably narrow.

This is directly relevant for the Chilean case. The observed male density 
($58.3\%$) falls below the lower bound of the feasible region---meaning that under the baseline assumptions, no successful glidepath exists for the average Chilean male worker. The observed female density ($49.6\%$) lies well outside it. At current Chilean densities, the glidepath design problem is simply infeasible for the representative worker.

A critical implication is that portfolio investment decisions are not a 
substitute for adequate contribution density. If density cannot be raised 
materially above $58\%$, other levers---such as the contribution rate, the retirement age, or the replacement rate target---must be reconsidered, since no glidepath can deliver the required return ($R^*$) under the baseline assumptions at those density levels.

\paragraph{Gender gap.}

The preceding analysis relied on a gender-neutral representative worker. 
In reality, three key parameters differ sharply between men and women in 
Chile, all pushing $R^*$ higher for women: (i) earlier retirement age 
(60 vs.\ 65 years); (ii) longer post-retirement life expectancy (90 vs.\ 86 years); and (iii) lower contribution density ($49.6\%$ vs.\ $58.3\%$) 
\cite{dl3500,sp2023mortality,sp2024ficha}. Consequently, the gender-specific required returns differ substantially: $R^* = 5.3\%$ for men and 
$R^* = 8.4\%$ for women (Table~\ref{tab:gender}).

\begin{table}[H]
  \centering
  \caption{Gender-specific baseline simulation parameters.}
  \label{tab:gender}
  \begin{tabular}{@{}lcc@{}}
    \toprule
    Parameter & Men & Women \\
    \midrule
    Retirement age (years)          & 65      & 60      \\
    Life expectancy at retirement (years) & 86 & 90     \\
    Contribution density            & 58.3\%  & 49.6\%  \\
    Required return $R^*$           & 5.3\%   & 8.4\%   \\
    \bottomrule
  \end{tabular}
\end{table}

The case of women sits well outside the feasible region: contribution density 
alone places it below the $58\%$ lower bound identified in the previous 
subsection, and the other two factors---earlier retirement age and longer 
life expectancy---compound the problem. The implication is clear: no 
glidepath can deliver an adequate pension for the representative female 
worker under current conditions. Structural reforms are necessary 
complements---such as raising the retirement age for women, increasing 
contribution rates, or strengthening solidarity mechanisms (e.g., state-funded pension complements).

\section{Discussion of Results, Implementation, and Statistical Considerations}

Broadly speaking, it is not difficult to find a CVaR glidepath that appears 
attractive under a particular set of assumptions. A sufficiently rich grid of 
risk limits, together with sufficiently favorable assumptions about wages, 
contribution density, annuity pricing, asset selection, and future returns, 
can easily produce a glidepath that clears the chosen return target. The more 
relevant question is therefore not whether one successful glidepath can be 
found, but what the overall pattern of results says about the role 
of portfolio design in the Chilean TDF system.

With that in mind, a careful reader may reasonably ask the following questions. 
First, what exactly does $\Psi$ measure, and how should it be interpreted? 
Second, which findings are robust enough to serve as inputs for professional 
and regulatory discussion? Third, how should the terminal CVaR limit $B$ be 
interpreted in light of post-retirement risk and the choice of pension modality? 
Fourth, what role does the chosen asset universe play in the results? 
Finally, what statistical and modeling limitations should temper the numerical 
interpretation of the findings?

The following considerations clarify what the results mean, what they do not 
mean, and how far they can reasonably be taken as inputs for regulatory 
discussion.

\subsection{Core Interpretation and Inputs for Policy Discussion}

The technical interpretation of the results, and their relevance for regulatory discussion, proceeds in layers. We first clarify what the success statistic measures, then discuss how the glidepath results can inform regulatory design, and finally identify the economic constraints that can make the target return infeasible regardless of the glidepath selected.

With respect to the interpretation of $\Psi$, it is important to emphasize that $\Psi$ is not a probability of success for a particular manager following a fully specified investment rule. Formally, the empirical $\Psi$ defined in equation~\eqref{eq:psi-estimator} averages over two sources of randomness: the feasible portfolio trajectories assembled from monthly asset allocations sampled with Hit-and-Run, and the simulated returns scenarios contained in the returns  array $\mathcal{R}=\{R_{s,t}\}$. Thus, $\Psi$ should be read as the fraction of feasible portfolio--returns scenario combinations in which the annualized return clears $\Rstar$, under the risk constraints imposed by the glidepath. It is neither a best-case nor a worst-case statistic over the feasible set. Rather, it is an average feasibility measure: it asks whether the risk budget specified by the glidepath ($L(t)$ curve) leaves a sufficiently large mass of admissible  asset allocation trajectories capable of reaching the target return.

Given this interpretation, the central design problem is one of constrained 
optimality: among all CVaR glidepaths that clear the $\Psi > 50\%$ success 
threshold, how do different glidepaths trade off the probability of delivering a return higher than $R^*$ against cumulative risk $\Gamma$? The exercise operationalizes 
this question over a class of piecewise-linear CVaR curves. This class is 
deliberately simple, but not narrow in a regulatory sense: many relevant 
decreasing CVaR glidepaths can be represented directly in this form, and more 
irregular decreasing paths can be conservatively approximated by a 
piecewise-linear CVaR ceiling.

Within this class, and fixing the terminal CVaR limit at $B=3\%$---a 
reasonably conservative level for portfolios close to retirement---the 
minimum-risk optimum $(A,B,T_A)=(6\%,3\%,58)$ is the lowest-$\Gamma$ 
glidepath that clears the $\Psi > 50\%$ success threshold within the grid 
considered here. It should not be interpreted as a globally optimal solution 
over all possible risk policies, but as a meaningful benchmark conditional 
on the modeling choices of this exercise: the terminal CVaR limit of $3\%$, 
the selected asset universe, the return-generation engine, and the Monte 
Carlo implementation.

Beyond the exact optimum (however defined), the analysis shows that direct 
control of risk through portfolio-level CVaR metrics yields glidepath rules 
that are straightforward to interpret from a regulatory standpoint. Instead 
of prescribing age-dependent equity allocations, a regulator can evaluate 
admissible CVaR paths and ask whether they are compatible with delivering 
the required return $R^*$. The relevant trade-off is clear: lower early-life 
risk requires delaying the transition age $T_A$, while higher early-life risk 
allows de-risking to begin earlier.

A concrete illustration: what is the earliest age at which risk can begin to 
decline while still clearing the $\Psi > 50\%$ success threshold? Under a 
high early-life CVaR limit $A=10\%$ and the same terminal limit $B=3\%$ used 
throughout the analysis, the exercise identifies $T_A=45$ years as that boundary 
age. Glidepaths with $T_A < 45$ years fail to deliver $R^*$ even under this 
generous early-risk allowance, regardless of how much risk is taken during 
the constant-risk period. This illustrates the practical value of the 
framework: it translates a concrete policy question---how early can de-risking 
begin without compromising the ability to deliver adequate pensions?---into 
a clear, simulation-based answer.

The main economic constraint behind these  results is the contribution density. It is not merely another input parameter. It is the dominant factor in the Chilean application. The density exercise shows a narrow transition region: below roughly 59\%, no glidepath in the grid succeeds, while above roughly 67\%, virtually all do. This sharp transition is economically intuitive. Lower density reduces contributions, which raises $\Rstar$ nonlinearly because the worker has less capital to compound over the same horizon. But the way we model density is intentionally simple: a worker contributes a constant fraction $\beta$ of the statutory contribution every month. This preserves total career contributions but suppresses timing risk. In reality, missed contributions early in life are more damaging than missed contributions near retirement, because they forgo decades of compounding. Conversely, contribution gaps clustered late in life have a less damaging  effect on $\Kstar$ and on realized wealth. A richer model would simulate employment histories, unemployment spells, informality, maternity-related gaps, and wage scarring jointly. The qualitative conclusion should therefore be read conditionally but clearly: under the assumptions of the exercise, sufficiently low contribution density can make the glidepath problem infeasible.

The results for women highlight the same point. They should be understood as 
a diagnosis of structural differences, not as a suggestion that gender 
inequality in pensions can be addressed through portfolio investment decisions alone. 
The case of women is difficult because these three parameters all move in the 
same direction---lower contribution density, earlier retirement age, and 
longer life expectancy---each independently pushing $R^*$ higher. The 
combined effect is an $R^*$ so high that the investment problem becomes 
infeasible within the parameters  considered here. This is not, of course, a failure of the CVaR-glidepath methodology.

The results consistently show that allowing sufficient risk early in the 
working life is valuable, and that de-risking too early is costly. But in 
the case of women, even the most aggressive glidepath cannot deliver $R^*$ 
under current conditions.  Structural reforms such 
as raising the retirement age for women, increasing contribution rates, or 
strengthening solidarity mechanisms are required.

\subsection{Choice of the Eligible Assets and Implications}

A further consideration concerns the specification of the eligible asset universe. The objective of the exercise is not to identify the best possible collection of indices or investment strategies, nor to solve a portfolio-optimization problem over the full set of instruments available to Chilean pension managers. Rather, this study  uses a compact and economically plausible set of indices spanning the main asset classes relevant for the Chilean pension system: local fixed income, local equities,  global fixed income, global equities, and alternative assets. The eligible universe was treated as an input specification--- its composition was not the object of any search, and we did not iterate over alternative index sets in order to maximize $\Psi$. Likewise, the asset universe was not tuned \textit{ex post} to improve the reported results. This is important because the purpose of the exercise is to evaluate whether a given CVaR-limit glidepath leaves a sufficiently large set of feasible portfolio trajectories capable of reaching the target return, not to overfit the investment universe itself.

This distinction matters for the interpretation of the results. A richer asset universe, or a more carefully selected set of active strategies, could improve the options available to an optimizing manager, and in that sense the current universe should not be read as an exhaustive opportunity set. However, $\Psi$ is not associated with an optimized portfolio---it is an average  measure over sampled feasible portfolios--returns scenarios combinations. Expanding the asset universe can therefore shift the results in either direction: adding high-quality return sources may improve $\Psi$, while adding low-return, redundant (highly correlated)  or poorly compensated assets may lower it. The exercise should therefore be read as a test of whether a representative, but not exhaustive, investment universe is capable of supporting glidepaths that achieve $\Psi > 50\%$.

\subsection{Diversification and Liquidity Considerations}

We note that the baseline exercise does not impose an explicit diversification constraint, such as a maximum weight per asset class or a minimum effective number of holdings, nor does it impose an explicit liquidity constraint. This modeling choice reflects the objective of the paper: to evaluate CVaR-limit glidepaths, not to prescribe a specific asset allocation rule or a complete investment mandate. In practice, diversification and liquidity requirements could be added as additional admissibility constraints without altering the structure of the framework.

This omission does not invalidate the main results for three reasons. First, the statistic reported in the paper is not based on a single optimized portfolio. It is computed over randomly sampled feasible allocations, so the evaluation reflects the behavior of a broad set of admissible portfolios rather than the performance of an extreme corner solution. In this sense, the exercise is less exposed to the concentration problems that often arise in unconstrained portfolio optimization.

Consistent with this interpretation, as a diagnostic check, we compute the Herfindahl--Hirschman Index (HHI) of the sampled feasible allocations,
\[
\mathrm{HHI}(\Omega)=\sum_{j=1}^{N}\omega_j^2,
\]
where $\omega_j$ is the portfolio weight of asset $j$ and $N$ is the number of eligible indices. Lower values indicate more diversified allocations; with nine eligible indices, the equal-weight lower bound is $1/9 \approx 0.111$. For the five successful boundary glidepaths reported in Table~\ref{tab:boundary}, the HHI is remarkably stable across curves. Across 4.8 million allocation-month observations per curve, the mean HHI ranges from 0.1987 to 0.1996, the median ranges from 0.1885 to 0.1888, and the 90th percentile remains between 0.259 and 0.261. In effective-number terms, the typical sampled portfolio is therefore equivalent to roughly five equally weighted positions (since 1/0.19 $\thickapprox$ 5). This evidence suggests that the randomly sampled feasible allocations are not dominated by highly concentrated corner portfolios.

Second, the eligible assets are represented by broad market indices. These indices serve as proxies for diversified investment vehicles, and most correspond to liquid public-market exposures. The relevant unit of allocation in the exercise is therefore not an individual security, but an already diversified investment strategy.

Third, in the particular Chilean calibration case studied here, the terminal CVaR limit indirectly restricts the use of high-risk alternative-asset proxies near retirement. Because the glidepath converges to a low terminal CVaR limit, terminal portfolios naturally shift toward lower-risk public-market exposures and away from asset classes whose risk profiles are inconsistent with that constraint. In this calibration, those lower-risk exposures also tend to be the more liquid public-market components of the eligible universe. This should not be interpreted as a substitute for a formal liquidity rule: if policymakers wish to guarantee a minimum level of liquid assets in terminal portfolios, such a requirement should be imposed explicitly.

\subsection{Some Considerations Regarding Potential Statistical Biases}

Several parameters were fixed by design before the simulation, including the
minimum-sufficiency target ($\Psi > 50\%$), the $90\%$ CVaR confidence level,
the terminal CVaR limit, and the piecewise-linear glidepath structure. We then
applied the optimality criterion of the paper: among the glidepaths satisfying
the sufficiency constraint ($\Psi > 50\%$), we select the one with the lowest
cumulative risk. This limits overfitting because the risk metric and policy class
are not tuned \textit{ex post} to improve the results. The optimum is therefore conditional
on these modeling choices, but it is not the product of searching over them.

The return-generation model is another possible source of statistical bias.
Financial prices are not generated by an independent Gaussian process: returns
can exhibit regime changes, volatility clustering, time-varying correlations,
and asymmetric tail dependence. The Gaussian-copula engine is useful because
it preserves empirical marginal behavior and is computationally tractable, but
it is still a simplification of a more complex time-series process. We do not
view the construction of a superior return generator as the goal of this
study. The framework is modular, and alternative engines---such as block
bootstraps, regime-switching models, $t$-copulas, synthetic data, or stressed
scenarios---could be employed in conjunction with the framework presented here.

Finally, the CVaR estimates and the Hit-and-Run samples are finite-simulation objects. Each monthly CVaR uses the worst 10\% of $S=10{,}000$ simulated return outcomes, and the feasible asset allocations are sampled with a burn-in of $b_{\mathrm{burn}}=20$. In practice, the impact of this short burn-in is mitigated by the scale of the retained sample: for each month we keep $I=10{,}000$ post-burn-in draws, and the cross-month permutations described in the Appendix    further dissociate the trajectories used to evaluate $\Psi$ from the sequential ordering of any single Hit-and-Run chain. These implementation choices are therefore unlikely to drive the broader patterns reported above, which are dominated by the shape of the glidepath  and contribution density.

\section{Concluding Remarks}

The following remarks summarize the main contributions and findings 
of this study. They are organized in two groups: the first addresses 
the proposed methodology and its broader applicability; the second 
covers some substantive conclusions related to  the Chilean pension reform.

\paragraph{Methodological Observations.}

\begin{itemize}
\item The framework presented herein is an effective tool for TDF design when a clear 
return objective can be specified. By translating that objective into 
a minimum required return $R^*$, it identifies multiple candidate 
glidepaths---each defined by a declining CVaR constraint---that are 
capable of delivering $R^*$ with a satisfactory probability. Moreover, 
the framework is agnostic to the choice of return-generation engine: 
alternative models---such as block bootstraps or regime-switching 
specifications---can be substituted without altering the core 
structure.

\item The Hit-and-Run algorithm provides an efficient 
mechanism for sampling uniformly over the feasible allocation space 
at each point in time. Although the technical details are confined 
to the Appendix, it is worth emphasizing that the algorithm is 
general: it applies to any convex feasible set defined by a 
portfolio-level risk constraint, regardless of the number of assets 
or the specific form of the CVaR limit curve.

\item The fact that the method is able to identify multiple successful 
glidepaths---those clearing the $\Psi > 50\%$ threshold---naturally 
invites the incorporation of additional selection criteria. The cumulative 
risk metric $\Gamma$ introduced in this paper provides one such criterion, 
allowing the selection of the least risky glidepath among all successful 
candidates. Other potential criteria could include a minimum diversification 
requirement (e.g., measured via the Herfindahl-Hirschman index), a maximum 
concentration limit by asset class, or a maximum drawdown constraint. 
The framework can accommodate such extensions without modification to 
its core structure.

\item The method decomposes the portfolio selection  process into 
two distinct steps. A risk manager first defines, at each point in 
time, the set of feasible asset allocations consistent with the 
CVaR glidepath. (It is implicitly assumed that the glidepath, presumably specified by the regulator, is consistent with a reasonably  high probability of achieving the desired return.)   Then,  the  portfolio manager  selects from within that 
set  an asset allocation. This formulation places less reliance on the manager's ability 
to identify an optimal portfolio, since he/she is making the selection from a set of choices that by design   carry a high probability of meeting the 
return target.

\item At a broader level, the approach presented here offers a complement to the 
canonical (Markowitz) mean--variance paradigm and to later risk-based portfolio-optimization methods, including CVaR optimization in the sense of Rockafellar--Uryasev \cite{rockafellar2000}. Rather than selecting a single optimal portfolio by maximizing return 
subject to a risk constraint, the focus here is on identifying a 
sufficiently large set of portfolios capable of delivering a return 
above a minimum threshold with high probability. The distinction 
is subtle but meaningful: it reframes the portfolio selection problem as  a
feasibility-and-sufficiency problem, rather than an optimization 
one, and invites further reflection on the role of the portfolio 
manager in the investment process.

\end{itemize}

\paragraph{Chilean Pension Reform Observations.}

\begin{itemize}

\item Under reasonable assumptions---specifically, a contribution 
rate of $16\%$ and a contribution density at or above $60\%$---the 
framework confirms that it is indeed possible to design TDFs capable 
of delivering adequate pensions. In fact, a substantial number of 
candidate glidepaths clear the $\Psi > 50\%$ success threshold, 
suggesting that the design problem is feasible under realistic 
conditions and that this conclusion is robust rather than the product 
of a single favorable choice of simulation parameters.

\item Two parameters emerge as the most important factors that drive the success potential of a glidepath:  (i) the transition age $T_A$---the age at which the fund 
begins reducing risk---is a critical design variable; glidepaths that de-risk too early consistently fail to deliver the required return ($R^*$), 
regardless of how much risk is taken in the early phase; and (ii) contribution 
density acts as a hard constraint---below a critical threshold, no 
glidepath succeeds, and portfolio design alone cannot compensate for 
structurally low contribution rates. These two findings together 
define the boundary conditions within which TDF design is possible.

\item The analysis highlights a stark gender gap. Under current 
conditions---earlier retirement age, longer life expectancy, and 
lower contribution density---the required return $R^*$ for women 
lies well outside the feasible region, meaning that no glidepath 
can deliver an adequate pension for the representative female worker. 
This is not a limitation of the framework; it is a diagnosis. 
Adequate pensions for women may therefore require complementary policy measures, such as higher contribution rates, adjusted 
retirement ages, or strengthened solidarity mechanisms.

\item A key methodological lesson from Chile's own experience with 
the \textit{multifondos} system is directly relevant here. When risk 
is controlled indirectly---through asset-class limits rather than 
at the portfolio level---funds do not necessarily rank by risk 
as intended: the most conservative fund may, over extended periods, 
outperform the riskiest one, and Sharpe-ratio rankings can be 
systematically inverted relative to the regulator's original intent 
\cite{schlechter2019,pagnoncelli2023}. The \textit{fondos 
generacionales} are exposed to the same risk if portfolio-level 
risk management is not embedded in their design. A TDF intended 
for workers aged 55, for example, should, by construction, carry less risk than 
one intended for workers aged 25---at all times, not just on 
average. Direct control of risk via a portfolio-level CVaR 
constraint, as proposed here, provides that guarantee in a way 
that asset-class limits alone cannot.

\item The framework is well suited to inform the regulatory discussion 
surrounding Chile's upcoming pension reform. By making the trade-offs 
between risk, return, and contribution density explicit and 
quantifiable,  the method presented provides policymakers with a transparent and 
flexible tool for evaluating competing glidepath (TDF) designs.  
\end{itemize}

\bibliographystyle{unsrtnat}
\bibliography{references}

\clearpage
\phantomsection
\addcontentsline{toc}{section}{Appendix}
\section*{APPENDIX\quad Uniform Sampling over the Feasible Set via the Hit-and-Run Algorithm}
\label{app:hit-and-run}
\renewcommand{\thesubsection}{\arabic{subsection}}
\renewcommand{\thesubsubsection}{\arabic{subsection}.\arabic{subsubsection}}
\setcounter{subsection}{0}
\subsection{The Feasible Set}

As established in the Computational Strategy section, for each month $t \in \{1,\ldots,Q\}$ we need to construct a set of $I$ feasible asset allocations. An allocation $\Omega=(\omega_1,\ldots,\omega_N)^{\top}$ is feasible for month $t$ if it satisfies three conditions simultaneously: the weights are non-negative ($\omega_i \geq 0$ for all $i$), they sum to one, and the estimated CVaR of the portfolio does not exceed the limit $L(t)$ set by the glidepath.

Formally, the feasible set for month $t$ is defined as:
\begin{equation}
F(t)=\left\{
\Omega \in \mathbb{R}^{N}:
\omega_i \geq 0\ \forall i,\quad
\sum_{i=1}^{N}\omega_i=1,\quad
\cvar(\Omega,t)\leq L(t)
\right\},
\end{equation}
where the estimated CVaR is computed from the $S$ scalar portfolio returns under the scenarios for month $t$,
\begin{equation}
\delta_{s,t} = \Omega^{\top}R_{s,t}, \qquad s=1,\ldots,S,
\end{equation}
by sorting them from lowest to highest and averaging the worst 10\%, since we work with a 90\% confidence level.

Geometrically, $F(t)$ is the intersection of the standard $(N-1)$-simplex with the sublevel set $\{\Omega:\cvar(\Omega,t)\leq L(t)\}$. This latter region has no closed analytical form, since the estimated CVaR is a piecewise linear function of $\Omega$, giving $F(t)$ a convex but irregular geometry. Since direct sampling from $F(t)$ is not straightforward, we use the Hit-and-Run algorithm, a Markov Chain Monte Carlo (MCMC) method designed to sample uniformly over convex bounded sets.

\subsection{The Hit-and-Run Algorithm}

Starting from a feasible point, at each iteration the algorithm picks a random direction, computes the chord of the line through the current point that stays inside the feasible set, and jumps to a point drawn uniformly along that chord. Under standard regularity conditions, the stationary distribution of the chain is the uniform distribution over $F(t)$. We describe each component of the algorithm as implemented.

\subsubsection{Initial Feasible Point}

Initializing the chain requires a point $\Omega^{(0)} \in F(t)$. We attempt the following in order: (i) the equal-weight portfolio $\Omega=(1/N,\ldots,1/N)^{\top}$; (ii) portfolios drawn at random over the simplex; and (iii) if neither of the above satisfies the CVaR constraint, we solve an optimization problem that minimizes the CVaR of $\Omega$ over the simplex using the SLSQP method, and verify that the minimum found is feasible. If the problem is feasible, which in practice always holds if $L(t)$ is not excessively restrictive, $\Omega^{(0)}$ is obtained.

\subsubsection{Direction Generation and Projection onto the Unit-Sum Hyperplane}

Given the current point $\Omega^{(k)}$, a random direction $d$ is generated by drawing $z \sim N(0,I_N)$ and normalizing:
\begin{equation}
d=\frac{z}{\lVert z\rVert}.
\end{equation}
This direction is then projected onto the hyperplane $\{v \in \mathbb{R}^{N}:\sum_i v_i=0\}$ via:
\begin{equation}
d \leftarrow d - \frac{1}{N}\sum_{i=1}^{N}d_i \cdot \mathbf{1},
\end{equation}
and renormalized. This projection is essential: it ensures that moving along $d$ does not alter the sum of the weights, thereby preserving the constraint $\sum_i \omega_i=1$ at every subsequent step.

\subsubsection{Computing the Feasible Line Segment}

The next step is to determine the interval $[t_{\min},t_{\max}]$ of scalar values $\tau$ such that $\Omega^{(k)}+\tau d$ remains in $F(t)$. This interval is computed by combining two sources of constraint.

\paragraph{Non-negativity constraints.}
For each asset $i$ with $d_i \neq 0$, the value of $\tau$ that drives $\omega_i^{(k)}+\tau d_i$ to zero is:
\begin{equation}
\tau_i^{*}=-\frac{\omega_i^{(k)}}{d_i}.
\end{equation}
If $d_i>0$, this yields a lower bound on $\tau$; if $d_i<0$, an upper bound. Taking the maximum of the lower bounds and the minimum of the upper bounds yields the interval $[t_{\min}^{\mathrm{lin}},t_{\max}^{\mathrm{lin}}]$ consistent with non-negativity.

\paragraph{CVaR constraint.}
Within $[t_{\min}^{\mathrm{lin}},t_{\max}^{\mathrm{lin}}]$, the CVaR is a convex function of $\tau$. An adaptive search is carried out from each endpoint of the interval inward to identify the values $t_{\min}^{\cvar}$ and $t_{\max}^{\cvar}$ beyond which $\cvar(\Omega^{(k)}+\tau d,t)>L(t)$. The final feasible segment is:
\begin{equation}
\left[t_{\min},t_{\max}\right]
=
\left[
t_{\min}^{\mathrm{lin}} \vee t_{\min}^{\cvar},
t_{\max}^{\mathrm{lin}} \wedge t_{\max}^{\cvar}
\right].
\end{equation}

\subsubsection{Sampling and Update}

We draw $\tau^{*}\sim\mathrm{Uniform}(t_{\min},t_{\max})$ and set:
\begin{equation}
\Omega^{(k+1)}=\Omega^{(k)}+\tau^{*}d.
\end{equation}
A component-wise clip to $[0,1]$ and renormalization are applied to correct for floating-point rounding errors. If the resulting point satisfies all constraints, it is accepted as the new chain state; otherwise, the chain remains at $\Omega^{(k)}$.

\subsubsection{Burn-in and Sample Collection}

The first $b_{\mathrm{burn}}$ iterations of the chain are discarded (burn-in), as these early draws are still correlated with the starting point $\Omega^{(0)}$ and are not representative of the stationary distribution. We use $b_{\mathrm{burn}}=20$ in our implementation. After burn-in, the subsequent $I$ points are stored as the $I$ feasible allocations for month $t$.

An independent chain is run for each month $t$, so allocations from different months share no internal MCMC state.

\subsection{Handling Serial Correlation Across Months}

As noted in the Computational Strategy section, Hit-and-Run generates the $I$ allocations for each month sequentially. Although the marginal distribution of each draw converges to the uniform distribution over $F(t)$, adjacent draws within a given month exhibit serial correlation, as is inherent in any Markov chain.

If one were to use row $i$ of the $I \times Q$ allocation array directly, that is, the $i$th draw produced by Hit-and-Run in each month, the monthly allocations of portfolio $i$ would not be independent across months. To remove this artificial dependence, we apply an independent random permutation within each column of the array before extracting rows. As a result, the allocation of portfolio $i$ in month $t$ corresponds to a draw selected at random from the $I$ feasible allocations for that month, with no relation to its position in the MCMC sequence.

\end{document}